# *OPIMA:* Optical Processing-In-Memory for Convolutional Neural Network Acceleration

Febin Sunny, Amin Shafiee, Abhishek Balasubramaniam, Mahdi Nikdast, Sudeep Pasricha
Electrical and Computer Engineering Department
Colorado State University, Fort Collins, Colorado, USA
{febinps, amin.shafiee, abhishek.balasubramaniam, mahdi.nikdast, sudeep}@colostate.edu

*Abstract*—Recent advances in machine learning (ML) have spotlighted the pressing need for computing architectures that bridge the gap between memory bandwidth and processing power. The advent of deep neural networks has pushed traditional Von Neumann architectures to their limits due to the high latency and energy consumption costs associated with data movement between the processor and memory for these workloads. One of the solutions to overcome this bottleneck is to perform computation within the main memory through processing-in-memory (PIM), thereby limiting data movement and the costs associated with it. However, DRAM-based PIM struggles to achieve high throughput and energy efficiency due to internal data movement bottlenecks and the need for frequent refresh operations. In this work, we introduce *OPIMA*, a PIM-based ML accelerator, architected within an optical main memory. *OPIMA* has been designed to leverage the inherent massive parallelism within main memory while performing high-speed, low-energy optical computation to accelerate ML models based on convolutional neural networks. We present a comprehensive analysis of *OPIMA* to guide design choices and operational mechanisms. Additionally, we evaluate the performance and energy consumption of *OPIMA*, comparing it with conventional electronic computing systems and emerging photonic PIM architectures. The experimental results show that *OPIMA* can achieve 2.98× higher throughput and 137× better energy efficiency than the best-known prior work.

## I. INTRODUCTION

For emerging machine learning (ML) models being used across application domains [1]-[3], the exponential growth in their computational demands has significantly outpaced the rate of advances in traditional computing architectures [4], [5]. The resulting "Von Neumann bottleneck" that alludes to the memory wall problem [6], is a critical challenge to overcome, to support modern ML workloads. In response to the limitations posed by the Von Neumann architecture, various alternative paradigms are being explored by industry and academia. A promising alternate computing paradigm involves in-memory computing or processing-in-memory (PIM) [7]. PIM architectures propose a departure from traditional designs by integrating processing capabilities within the memory subsystem. This integration aims to minimize data movement, reduce latency, and minimize energy consumption associated with processing applications.

Given that dynamic random-access memory (DRAM) is the standard main memory technology today, it is an obvious candidate for PIM. Several prior efforts have focused on architecting DRAM-PIM [8]-[10]. However, conventional DRAM-based PIM systems have encountered challenges in achieving high throughput and energy efficiency. These challenges arise primarily due to internal data movement bottlenecks and the necessity for frequent memory refreshes. To address the energy and latency concerns associated with refreshes, non-volatile memory (NVM) technologies, such as ReRAM [11], [12], Spin-Transfer Torque RAM (STT-RAM) [13], and Phase Change Material (PCM) memories [14]-[16], have been considered. However, ReRAM and STT-RAM technologies face fabrication challenges and endurance issues [17], [18]. ReRAM additionally suffers from resistance drift over time, which impacts data readout accuracy [17].

PCMs offer better energy efficiency, bit density, and bandwidth than other NVMs. They can switch between two physical states: amorphous and crystalline. This switch results in a contrast in electrical resistance, allowing these materials to encode information based on varying resistance levels. In the context of electrically controlled PCM (EPCM) devices, these phase changes are induced by applying current through microheaters. It is possible to precisely regulate the phase shift from amorphous to crystalline, enabling the creation of multi-level cells (MLCs) to store more data by adjusting the extent of the material's crystallization. However, utilizing the resistance in PCMs to encode data poses challenges as the resistance values that PCMs attain depend non-linearly on the applied write voltage [19].

To address these challenges, optically programmed PCM (OPCM) cells can be considered [23]. OPCM cells are fabricated with PCM deposited on top of a photonic waveguide and are programmed through laser pulses. Here, in place of resistance, the refractive index of the PCM is the physical property used to represent data. OPCMs can be programmed using laser pulses guided to them through on-chip waveguides. This makes them ideally suited for integration onto silicon photonic platforms. OPCMs are based on silicon photonics, which is an emerging field that integrates photonic systems with electronics. This platform offers several advantages over traditional electronic circuits, including high throughput and low energy consumption, for specialized computation tasks [19]-[22]. Merging this computational capability with an OPCM main memory could allow for high-speed in-memory computation without the data movement and refresh bottlenecks seen in DRAM-PIM.

In this paper, we explore how to architect a photonic main memory, to enable ML acceleration through PIM. We utilize the OPCM-based main memory from [23] as the backbone for our architecture and make several changes to it to support PIM. We have named our photonic PIM architecture for ML acceleration, *OPIMA*. We use convolutional neural networks (CNNs) to showcase the effectiveness of *OPIMA* for ML inference acceleration. The proposed PIM architecture is characterized by multi-bit density per cell enabling multiply and accumulate (MAC) operations to be performed directly within memory. This capability along with architecture-level innovations allows *OPIMA* to outperform the state-of-the-art in terms of ML inference throughput and energy efficiency. In summary, the novel contributions in this paper include:

- Scattering and back reflection-aware OPCM cell design to maximize bit-density and minimize read errors per cell;
- Full system design of an OPCM-based PIM architecture that can operate as a main memory while performing PIM;
- Comprehensive comparison of operational efficiency of *OPIMA* with state-of-the-art accelerators.

## II. BACKGROUND AND RELATED WORK

Before we discuss our PIM architecture and associated techniques, we review some fundamentals and background on PCMs, OPCM main memories, and photonic computing.

### A. Phase Change Materials (PCMs)

PCMs possess the ability to shift between amorphous and crystalline states, depending on the level of thermal energy applied. This energy must be sufficient to alter the material's temperature to either its melting temperature ($T_i$; for transitioning to the amorphous state) or its crystallization temperature ($T_g$; for shifting to the crystalline state). Transitioning to the amorphous state consumes more energy because its required melting temperature exceeds the crystallization temperature. It should be noted that it is possible to induce partial phase changes within PCMs, creating intermediate states by converting only a fraction of the material to either state. These transitions can be initiated through electrical or optical means. Electrical heating can be provided through PN junctions whereas optically achieving phase changes requires a laser pulse, whose power and duration must be tailored to the material's specific transition energy needs. Common materials used for PCM applications include $Ge_2Sb_2Te_5$ (GST), $Ge_2Sb_2Se_4Te$ (GSST), and $Sb_2Se_3$ [24].

The change in a PCM phase brings with it a change in the electrical and optical properties of the material. PCM's states have different electrical resistances and different optical refractive indices. These differences in characteristics can be leveraged for data representation, including multi-bit data representation, enabling dense PCM-based memories and as discussed in this paper, PIM architectures.

For EPCM applications, the high-resistance amorphous state is used to represent a binary 0, and the low-resistance crystalline state is used to represent a binary 1. This non-volatile change in resistance allows the PCM cell to be paired with an access transistor to form a 1T1R EPCM memory cell and a corresponding memory array of these cells, as described in many prior works (e.g., [26]-[29]). However as discussed earlier, EPCM memories face many challenges, such as asymmetric and high write latencies [30], non-linear response to write voltages, and resistance drift.

OPCM memories rely on shifts in the material's refractive index to modulate optical transmission, enabling data storage and retrieval [24]. A deep understanding of a PCM's optical properties is crucial for the effective deployment of OPCM memories. A significant refractive index contrast, ensuring a clear distinction in optical transmission between phases, is vital for reducing optical signal losses and noise [25], which could otherwise lead to readout errors. Similar to the importance of resistance contrast in EPCM memories, a high refractive index contrast improves the signal-to-noise ratio (SNR) during data readout. This is extremely important not just from a data fidelity standpoint but also from a photonic PIM standpoint, as we must ensure error-free data readouts to ensure error-free calculations in the analog domain where photonic computations occur.

### B. OPCM Memory

A main memory architecture should have the ability to store large amounts of addressable data, which can be effectively retrieved and modified, whenever needed by the computing system. DRAMs achieve this by having row- and column-addressable memory cells, arranged into mats of cells, which in turn get organized into subarrays, and then banks. Collections of banks form memory chips, which are arranged into dual in-line memory modules (DIMMs) or 3D high bandwidth memory (HBM) architectures. Modern memory addressing schemes and memory controllers expect this style of data storage and management to be interfaced with them. So, it is prudent to consider a similar style of data storage with OPCM memory as well. A few recent works have tackled the challenge of building an addressable OPCM memory [23], [31], which can be used for the DRAM-like memory organization described above.

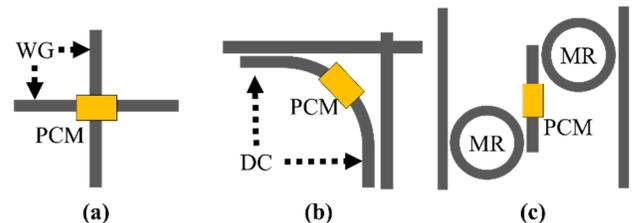

**Fig. 1:** OPCM memory cells proposed in (a) COSMOS [31]; (b) Photonic tensor core [15]; (c) COMET [23]. WG: Waveguide; DC: Directional Coupler; MR: Microring Resonator.

The work in [31] introduced a straightforward design for a crossbar-based cell, illustrated in Fig. 1(a), in which the OPCM is strategically positioned atop waveguide intersections. This cell design underpins the core of a main memory architecture called COSMOS. In this COSMOS OPCM memory, the mechanism for accessing data is facilitated by specific row and column access signals that operate on distinct optical wavelengths. These signals are required to be activated simultaneously to enable successful write operations within the memory structure. COSMOS also adopts a subtractive read

4technique. This method involves initially performing a read operation across an entire subarray. Subsequently, a reset signal is dispatched specifically to the row selected for reading, which clears its contents. Following this reset, the subarray undergoes another read operation. By executing this sequential reading and resetting process, it is possible to extract the data from the intended row. The two obtained readouts are subsequently processed through subtraction at the memory controller (MC). This intricate process, when combined with the assumption that each cell can store up to 4 bits of information, significantly amplifies the bit density achievable by this architecture, presenting a substantial advancement in memory design aimed at enhancing data storage efficiency and capacity. However, this architecture is inherently susceptible to optical crosstalk as the data storage mechanisms end up interfering with one another. It is especially susceptible to thermal crosstalk from write operations from adjacent rows, especially when multi-bit storage is assumed, as shown in [23].

The work in [15] showcased an OPCM cell, originally devised for photonic tensor core operation, but deserves discussion as it has been used in [32] for their OPCM memory-based ML acceleration work. The architecture has a crossbar structure to allow signals from orthogonal directions to interact with each other, enabling a wavelength-division multiplexing (WDM) based broadcast and weight computation technique [33]. The OPCM cell itself, however, is placed away from the waveguide crossing and can interact with a wavelength propagating along the horizontal waveguide. This interaction is enabled by the coupler on which the OPCM cell is fabricated. The coupler, as the name suggests, is a passive device designed to enable coupling between the WDM signal on the horizontal waveguide and the OPCM and does not offer wavelength selection like other photonic devices (e.g., a microring resonator (MR)). The work in [15] carefully designed this cell array to perform matrix-vector multiplications (MVM), where the matrix will be stored within the OPCM crossbar array, and individual elements of the vector are encoded per WDM signal batch in the horizontal waveguide. Each coupler has a splitting ratio associated with it, which is designed to ensure that the same fraction of signal from each wavelength reaches the GST OPCM, which is wavelength agnostic in the C-band of frequencies. So, in effect, each cell performs:

$$W_{cell} \times \left[\frac{\{A, \lambda_1\}}{n} + \frac{\{A, \lambda_2\}}{n} + \cdots + \frac{\{A, \lambda_n\}}{n}\right] = W_{cell} \times A \quad (1)$$

where, $W_{cell}$ is the weight stored in the OPCM, $A$ is the activation value imprinted onto the wavelength $\lambda_i$, $n$ is the WDM degree (i.e., the number of wavelengths in the WDM batch) that corresponds to the number of cells per row. This operation makes it an excellent MVM engine, with low latency and energy-efficient operation. Additionally, this cell (Fig. 1(b)) is compact and solves the interference and crosstalk issues that plague the COSMOS architecture [31] discussed earlier and would appear to be a good candidate for an OPCM-based PIM. However, the architecture is not column addressable, making it not a good choice for memory architecture. To consider this cell for a memory architecture and then a PIM architecture, column addressability to cells is essential.

To address these issues, the work in [23], COMET, designed a row and column addressable OPCM memory cell (Fig. 1(c)), which is also isolated from other cells to avoid data corruption due to crosstalk. This memory cell makes use of GST for data storage, with two MRs acting as the access control, electro-optically. The MRs are electrically tunable using a PN junction and are hence active when they are in resonance (turned on). In this active state, they allow signals propagating through the vertical waveguide on the left to access the OPCM cell. The data is imprinted onto the signal and is passed to the readout waveguide on the right (Fig. 1(c)). While the proposed cell is not as compact as the one suggested in [15], it offers more reliable data readouts, without crosstalk-induced errors. Further, the GST in the cell was designed to allow for improved energy efficiency in write operations. The subarray architecture also had provisions to ensure loss correction through intermittent semiconductor optical amplifier (SOA) arrays. There are several desirable characteristics that make COMET a suitable backbone for a PIM architecture, but there are also several challenges, as will be discussed in Section III.

### C. Photonic Computation

The previous subsection discussed the characteristics required to realize an OPCM main memory. In this subsection we discuss principles of photonic computation, which are a precursor to realizing a PIM solution with OPCM memory.

Photonic computation can be performed through either coherent or noncoherent (aka incoherent) analog computation methods [19]. Coherent photonic computation utilizes the phase of light waves in a controlled manner, enabling the encoding and manipulation (e.g., multiplication) of data via interference patterns. This approach takes advantage of the coherent properties of light, such as phase coherence and superposition, to perform complex mathematical operations rapidly and with high precision. Computing architectures that leverage coherent computing often make use of Mach-Zehnder interferometers (MZIs) for data manipulation through constructive or destructive interference with a single wavelength.

Noncoherent photonic computation, on the other hand, does not rely on the phase information of light, conventionally [33]. Instead, it involves manipulation of the intensity or amplitude of light waves to perform computations, making it less sensitive to phase fluctuations and coherence issues that might affect coherent systems. Noncoherent approaches are simpler in terms of data encoding and more robust as they do not have as many noise sources to deal with. This makes them suitable for a wide range of applications that require optical signal processing, such as image processing and sensor data analysis, and fundamental arithmetic operations. Additionally, they allow performing arithmetic operations at a very large scale, through the usage of WDM, making noncoherent photonics an attractive option for MVM and general matrix multiply (GEMM) operations. To leverage WDM signals, the photonic device used in noncoherent computation systems must be wavelength sensitive, which makes wavelength selective MRs popular candidates for the fundamental devices in these architectures.

An MR is an on-chip optical resonator, which resonates when it encounters an optical wavelength that matches its resonant wavelength ($\lambda_{MR}$). Through tuning mechanisms, $\lambda_{MR}$ can be altered, increasing losses to the encountered wavelength, thus enabling amplitude modulation, and hence forming the basis for noncoherent computation. There are two main tuning mechanisms used: thermo-optic tuning and electro-optic tuning. Both these mechanisms can change the effective refractive index ($n_{eff}$) of the bulk of the MR, thereby affecting ($\lambda_{MR} = 2\pi n_{eff} R$; $R$=MR radius). Thermo-optic (TO) tuning achieves this by heating the MR through microheaters, and electro-optic (EO) tuning achieves the same through free carrier injection via a PN junction fabricated across the MR [19].

Several noncoherent computation architecture in prior work [20]-[22] rely on MR operation for high throughput, reliable, low energy ML inference acceleration, through the computation technique called broadcast and weight (B&W) [33]. Here, MRs are tuned to reflect a stationary matrix, and vectors are introduced either as amplitude-modulated wavelengths or via a subsequent array of tunable MRs downstream from the initial MR array's output. The interaction of light with the MRs modifies its amplitude to reflect a multiplication operation. Several of these light signals can be summed using a photodetector, achieving $n$ multiply and accumulate (MAC) operations simultaneously. Here, $n$ is the WDM degree of the signal and should correspond to the size of the MR array.

From the discussions in Sections II.B, the OPCM memory cell in Fig. 1(c) is a potential candidate to be part of noncoherent architectures that perform computation operations. The OPCM cells can represent the stationary matrix/vector element, while the incoming light signal or one of the access control MRs can represent the changing vector. At this point, performing a memory read operation through the OPCM cell will achieve a multiplication operation. However, to achieve effective large-scale noncoherent computation via PIM, several challenges must be addressed, as discussed in the next Section.

## III. RE-ARCHITECTING *OPCM* MAIN MEMORY FOR PIM

In this section, we take a brief look at the COMET OPCM main memory architecture and why it cannot be used as is for effective noncoherent computation within a PIM solution.

The basic architectural component of the COMET main memory architecture is the OPCM memory cell depicted in Fig. 1(c). This memory cell is tiled to form an array, where each cell can be isolated from each other, while access is enabled through a wavelength assigned per column of the memory cells in the array. Row access is provided by turning on the access control MRs through EO tuning, thereby allowing the light signals access to the OPCM cell. $N \times N$ of these cells can form a subarray and $S \times S$ of these subarrays form a memory bank. A collection of $B$ memory banks constitute the main memory.

There are four major challenges that must be overcome to adapt the COMET OPCM memory architecture for PIM:

- Accessing all the cells in the same row across subarrays and banks requires $B \times S \times N$ wavelengths, which would be too energy- and power-expensive for a main memory of any reasonable size. During data read/write operations, the light signals are given access only to the subarray in which the corresponding row resides. This is achieved through the usage of GST-based waveguide switching, rather than splitting the WDM signal into multiple subarrays unnecessarily. It should be noted that using optical splitters and couplers would essentially multiply the laser power needed, and this must be avoided.

- COMET was architected with specific power consumption constraints, and hence many architectural choices were made to enable a power consumption of under 10W for the main memory operation, as discussed earlier. This power constraint allows it to operate in a similar power point to electronic main memory architectures such as DDR5. However, from a PIM perspective, these choices pose a problem. Having limited access to subarrays, and hence OPCM cells, per read/ write operation severely limits the achievable parallelization of computation operations. So, it is necessary to find a solution that enables multi-subarray access, without disrupting the optical main memory operation. Note that we cannot rely on increasing WDM degree or splitting signals from the source across multiple subarrays, as the power consumption this incurs will be many times higher than the 10 W constraint, reflecting the previous challenge.

- Optical signals can interact with each other in the readout waveguides. Increasing the WDM degree to avoid using splitters carries with it the risk of increased crosstalk and errors, especially when using OPCM cells at higher bit densities. So, careful orchestration of access and readout is necessary to achieve reliable and error-free computations.

- It is also important to consider the impact of bit density per cell on PIM operations. In COMET, a 4-bit per cell bit density was considered to ensure reliable memory operation. This limits possible neural network parameter sizes to 4-bit if there is a need to perform one-shot operations (e.g., multiplications) as discussed at the end of Section II. Without careful architectural considerations, it will be impossible to handle higher parameter sizes for computation within COMET.

In summary, there are several challenges associated with enabling PIM within an OPCM main memory. In our proposed *OPIMA* architecture, described in the next section, we address all these challenges via novel and significant alterations to an OPCM main memory architecture, to enable PIM within the memory platform, while still allowing it to retain its core functionality as a main memory solution.

## IV. *OPIMA* ARCHITECTURE

This section discusses the proposed *OPIMA* architecture and how it achieves PIM-based ML acceleration.

### A. Maximizing OPCM Memory Cell Efficiency

The *OPIMA* architecture is a PIM architecture that significantly expands the capabilities of the COMET main memory architecture. COMET explored how effective refractive index ($n_{eff}$) and optical absorption ($\kappa$) can be optimized for maximum energy efficiency in OPCM cells. Based on this analysis, the authors had selected GST as the best-suited OPCM material for the C-band of frequencies.

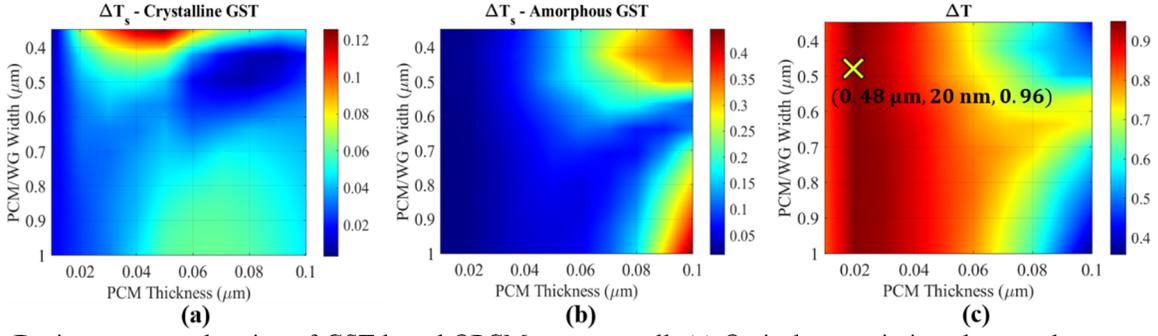

**Fig. 2:** Design-space exploration of GST-based OPCM memory cell, (a) Optical transmission changes due to scattering and back-reflections of the light ($\Delta T_s$) in the crystalline state, (b) $\Delta T_s$ in the amorphous state, (c) Optical transmission contrast between amorphous and crystalline states ($\Delta T$). Observe that for the chosen design point (highlighted with 'X'), the $\Delta T_s$ for both crystalline and amorphous states is less than 5% while the $\Delta T$ is at its maximum with 96%.

In this work, we consider more detailed factors influencing the behavior of OPCM-based memory cells, particularly the unwanted changes in the optical transmission of the cells because of the scattering and back-reflection of light when interacting with PCMs. The refractive index of the PCMs in crystalline and amorphous states is significantly higher than the refractive index of the waveguide material. Therefore, the propagating light can be scattered and reflected within the waveguide when interacting with the PCM on top of the waveguide. Such a scattering effect leads to unwanted optical transmission changes at the output of the OPCM memory cell.

To tackle this limitation, we performed a design-space exploration using GST on top of silicon waveguide to select the most optimal geometry that offers minimal transmission change due to light scattering and maximum transmission contrast due to phase change. To capture the optimal design with minimized scattering of the light, we use the following model:

$$T_{out} = T_{in} - \Delta T_s - P_{abs}, \quad (2)$$

where $T_{out}$ is the output transmission of the cell, $T_{in}$ is the input power, $\Delta T_s$ is the optical transmission change due to light scattering and back-reflections, and $P_{abs}$ is the total fraction of the power that is absorbed in the PCM cell (all in dB). We perform a design-space exploration of the PCM memory cell to minimize $\Delta T_s$ to minimize read errors stemming from the scattering effect of the light. For maximizing data signal strength, $\Delta T_s$ must be minimized so that the signal change due to written data ($P_{abs}$) is well represented in $T_{out}$:

$$T_{out} = (T_{in} - P_{abs}) \rightarrow \Delta T_s = 0. \quad (3)$$

This model is applicable to both amorphous and crystalline states of the cell. In addition, the desired OPCM memory cell should offer 1) high optical transmission which originates from the low power absorption in the amorphous state, and 2) high absorption and hence low optical transmission in the crystalline state. Consequently, the optimum design point should offer minimized light scattering and back-reflections at both crystalline and amorphous states while leveraging the high controlled optical transmission contrast. Therefore, the $\Delta T_s$ and the total optical transmission contrast between amorphous and crystalline states ($\Delta T = T_a - T_c$) can be used as a figure-of-merit to find the optimal design for the GST-based OPCM memory cell. This optimal design should offer a low $\Delta T_s$ in the amorphous and crystalline state and a high optical transmission contrast ($\Delta T$) between amorphous and crystalline states.

The design space exploration results for a 2-μm long GST cell that we designed are reported in Fig. 2. Observe that for the design point which offers the highest optical transmission contrast ($\Delta T$) highlighted in Fig. 2(c), the transmission changes due to light scattering and back-reflections is always less than 5% in the crystalline state (Fig. 2(a)) and the amorphous state (Fig. 2(b)). In addition, GST offers a high controlled optical transmission contrast (~96%) for the optimal design point shown in Fig. 2(c) which corresponds to a width of 0.48 μm and thickness of 20 nm. This higher contrast in transmission also allows us to program 16 levels of transmission per cell, allowing a bit density of 4 bits/cell.

The OPCM memory cell that we designed and optimized forms the building block of the *OPIMA* architecture that is designed for efficient data storage and access, as well as for performing in-situ multiplication operations. For the sake of maintaining row and column addressability, and hence main memory operation, we combine this OPCM memory cell with double MRs for optical access control.

*B. OPCM Memory Operation*

An overview of how *OPIMA* is designed to operate as a memory interfaced with an external general-purpose electronic CPU is shown in Fig. 3. A controller unit that handles the electro-optical interfacing requirements must reside between the CPU and *OPIMA*, as depicted in the figure. This controller unit interprets memory commands from the host CPU, enabling main memory operation. It also supports data caching for read data to be sent to the CPU or data to be written to the OPCM memory. In the latter case, the data is encoded via optical signals derived from the laser source.

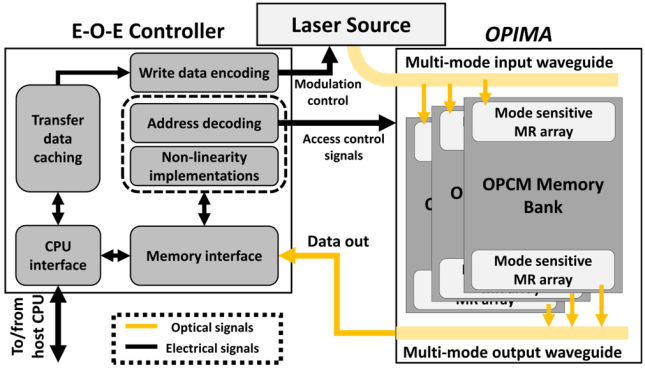

**Fig. 3:** Architectural overview of *OPIMA*

The isolated OPCM cells within *OPIMA* make read/write operations quite straightforward. For both operations, the row ID and subarray ID must be deciphered from the physical address. Once this has been done, laser signals are sent to the corresponding OPCM bank. The read process (Fig. 4(b)) happens as the signal passes through the memory cell and is modulated by the OPCM's optical transmission. The read data is sent back to the E-O-E controller where it is demodulated using an MR array. Then this data can be translated to the electronic domain and passed on to the CPU. The write process (Fig. 4(a)) requires much higher energy as it requires inducing partial phase transition in the OPCM memory cells. This necessitates more laser power to achieve the phase transition across multiple OPCM cells, based on the data to be written.

During the read and write operations, data integrity is a critical concern, especially considering the loss tolerance in signal transmission. *OPIMA* incorporates semiconductor optical amplifiers (SOAs) within and outside the banks and subarrays to maintain signal quality. We employ row-wise loss-aware signal amplification to counteract potential degradation. The banks and subarrays, once designed, have constant losses, facilitating this correction approach.

*C. OPIMA PIM Architecture*

As discussed earlier, the optical transmission of an OPCM cell modulates the optical signal passing through it. If the access control MR is tuned to represent the second parameter, the successive modulations from the MR and the OPCM can achieve a multiplication operation. However, since we need all the MRs in a row to behave identically to facilitate row access, it is better to tune the incoming laser signal to represent the second parameter. To achieve an accumulate operation, we must let two signals of the same wavelength, modulated to reflect products, interact with each other. To perform this step, we need to involve products from another subarray sharing the same readout waveguide bus. Within the readout waveguide bus, these signals interfering with each other generate the sums. This is desirable from a PIM perspective but will lead to erroneous readouts from a main memory perspective. Hence, for achieving this goal and thus realizing the PIM operations for ML inference acceleration, we need several architectural changes to the main memory architecture, as discussed next.

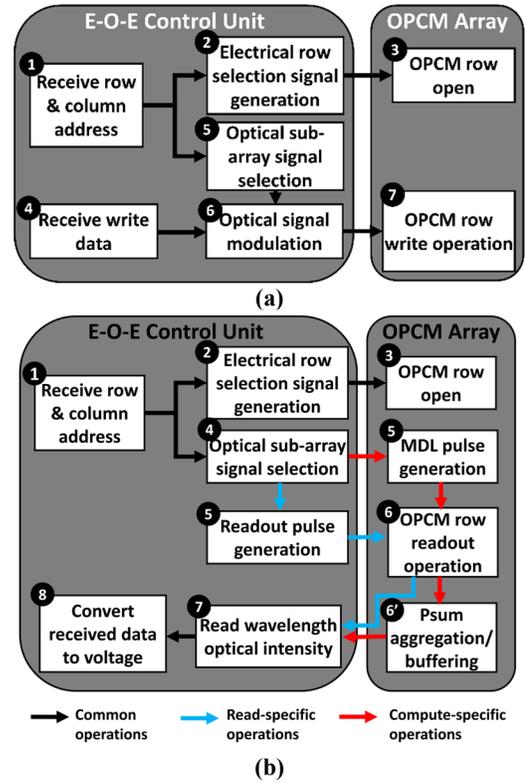

**Fig. 4:** Memory (a) write and (b) read operation in *OPIMA*; *OPIMA* utilizes multiple read signals simultaneously to perform computation operations. The differences in control flow between a memory read operation and performing in-memory computation are highlighted in (b).

To realize high throughput and error-free PIM operation in *OPIMA*, we need to address four major challenges: (1) We need to leverage additional mechanisms to increase memory access and computation parallelism beyond those offered by WDM; (2) reads should be supported from a selected subarray or a group of subarrays as needed, without interrupting the main memory operation; (3) When simultaneously read out, the data from computation outputs and main memory accesses must not interfere with each other in an undesirable manner; and (4) the architecture should support PIM operations between parameters (e.g., CNN weights and activations) of any size, irrespective of the specific bit density used in the OPCM cells.

*1) Implementing MDM for improved parallelism*

To address challenge (1), within *OPIMA*, we design the multi-bank OPCM memory organization to go beyond WDM and additionally use mode-division multiplexing (MDM) to enable parallel access across banks (Fig. 5(a)). MDM involves exciting higher order modes in a MDM waveguide bus, where each of the modes of a wavelength can then be used for supporting parallel data transfers and computations. Note that multiple wavelengths co-existing in the waveguide bus (WDM) provide further parallelism for data transfers and computations. Increasing the number of modes comes at the cost of increased width of the individual waveguide to allow the higher order modes to be excited and propagated, as well as increased crosstalk. Thus, determining the optimal number of modes (MDM degree) requires a careful trade-off analysis.

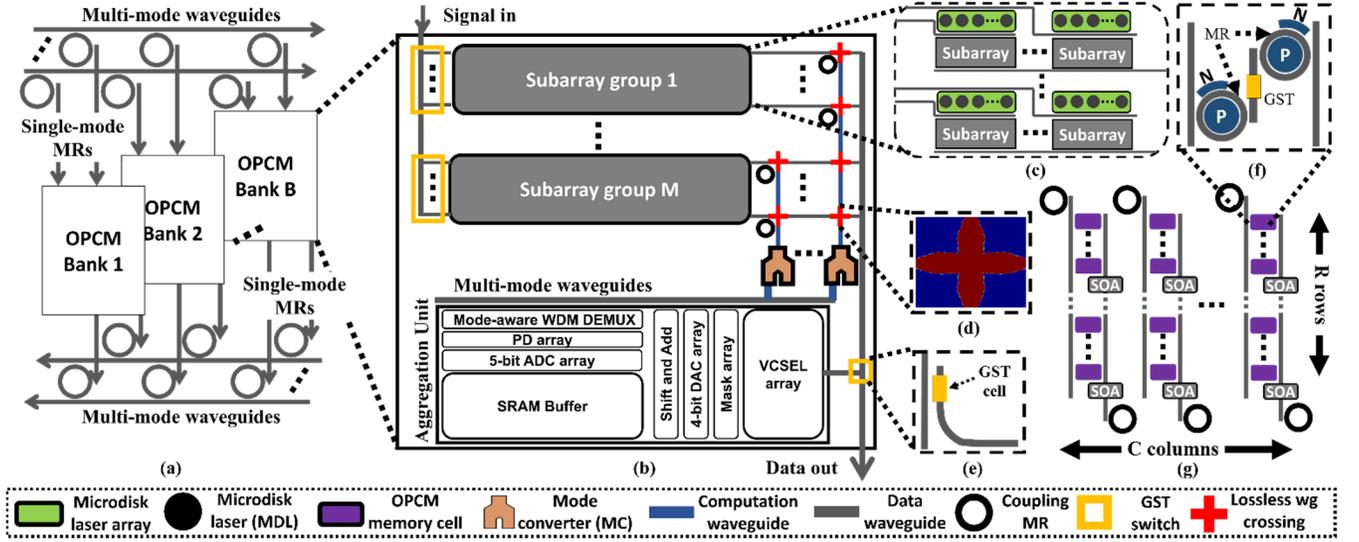

**Fig. 5:** *OPIMA*'s PIM-specific architecture; (a) OPCM bank organization; (b) Subarray organization within the bank, showcasing grouping, aggregation unit, and computation specific waveguides, coupling MRs, and mode converters (MC); (c) Subarray group internals; each subarray is equipped with a microdisk laser (MDL) array for PIM operation independent of main memory operation; (d) low loss waveguide (wg) crossings designed using inverse design; (e) GST cells used for subarray access control during OPCM main memory operation; (f) OPCM memory cell with EO tuned MRs showcased; (g) OPCM memory array within subarrays, with $R \times C$ OPCM cells within it.

We inverse designed photonic mode convertors based on [34] to exploit the first four modes of TE polarization. Compared to conventional mode convertors based on tapered structures or thickness changes to induce the required index changed, the inverse designed mode convertors offer a compact footprint and minimal loss. Note that exciting more than four modes in the waveguide at the same time is physically challenging as it requires extremely wide waveguides that significantly increase memory area. In addition, higher order modes suffer from intermodal crosstalk due to the overlap of the modes [35], [36]. Based on our MDM propagation analyses, we decided to keep the MDM degree to four, which limits the number of banks in the architecture to four. These MDM signals can be filtered by mode-sensitive MRs to their respective banks and be routed to their respective subarrays through GST switches, enabling parallel read/write operations across banks. However, there is a need to improve parallelism further to achieve higher PIM throughput. Additionally, while it is technically possible to perform a MAC operation by reading from two OPCM cells, this operation will be limited to 4-bit parameters under the configuration discussed here.

*2) Redesigning banks for concurrent PIM and memory access*

A memory bank within the *OPIMA* architecture is composed of $R \times C$ OPCM cells (Fig. 5(g)), offering a total capacity determined by the product of the number of cells and the bit density of each OPCM Multi-Level Cell (MLC). To enhance energy efficiency, banks are divided into subarrays. The *OPIMA* architecture employs electrically controlled GST-based waveguide switching to facilitate efficient subarray access (Fig. 5(e)), markedly reducing the laser power requirements. The GST switch introduces minimal losses and is pivotal for the energy-efficient operation of the system. We need to make changes to this organizational structure and provide additional access mechanisms to address challenge (2).

Data within OPCMs cannot be sensed in the same manner as charge-based storage in DRAM. Accessing data in OPCM cells necessitates external laser signals, which must overcome several losses in propagation, to be rerouted to the subarrays within which the OPCM cell resides. This leads to high power consumption, to overcome the losses and the signals being split into several destinations. To circumvent this, we propose the addition of local laser sources to subarrays, which can be triggered as needed for reads. Fortunately, unlike OPCM write operations, OPCM read operations are not energy intensive [23] and hence we can employ low-power lasers.

For *OPIMA* we opted for low-power microdisk laser (MDL) arrays (Fig. 5(c)), which can be integrated with every subarray. Each subarray uses $C$ MDLs in its subarray, reflecting the column number per subarray. The laser output from the MDL array can be coupled onto the signal input waveguide of the corresponding subarray, using directional couplers. Using the MDL arrays, we can access any row within a subarray, without the involvement of the external laser source which drives the main memory operation. Additionally, since the MDL arrays are independent of each other, multiple of them can be activated simultaneously to read from multiple subarrays without having to reroute or incur additional losses.

Moreover, to ensure that we can read for PIM while main memory operations happen in parallel, the subarrays are divided into several groups (Fig. 5(b)). One row of subarrays per group can be employed for PIM at a time, while the rest of the subarrays can be used for main memory read/write operations. This ensures significant parallelism in MAC operations that can be executed simultaneously per bank, offering simultaneous solutions to challenges (1) and (2).

*3) Reducing output interference*

Now that we have several MAC operations being supported simultaneously, we must ensure that their results can be aggregated without interfering with each other or the main memory readout operations, to address challenge (3). It should be noted that the subarrays make use of WDM signals which can interfere with each other constructively or destructively.

To avoid computation signals interfering with memory read operations, we employ a series of computation-specific waveguides. Computed data is rerouted to the computation waveguides rather than the data-out waveguide using coupling MRs which can be activated alongside the MDL array (Fig. 5(c)). The computation waveguide is used to move the data to the aggregation unit in the bank. To prevent losses and the computed signal from interfering with orthogonally traveling data signals, all the waveguide crossings in the computation waveguide have been carefully designed to be as leakage-free as possible (Fig. 5(d)).

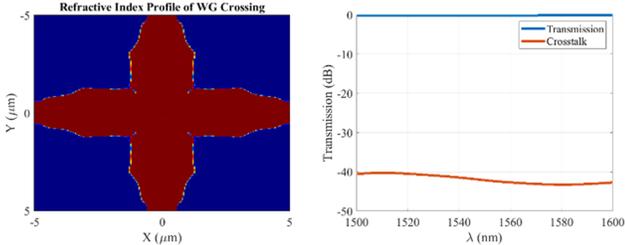

**Fig. 6:** Low-loss waveguide crossing designed with inverse design methodology (left) and its loss profile for C-band (right).

To achieve the optimized waveguide crossing design, we used a photonic inverse design technique to minimize the loss and crosstalk of the waveguide crossings. The Lumerical FDTD solver [37] with the LumOpt [38] inverse design library was used to perform the geometry optimization of the waveguide crossings. The optimized geometry of the waveguide crossing is shown in Fig. 6. Note that the transmission of the fundamental TE mode was used as a figure-of-merit in our inverse design optimization of waveguide crossing. We can observe from the figure that the inverse-designed waveguide crossing offers the maximum transmission at the C-band with less than 0.001% of the input optical signal being lost due to optical insertion loss. Note that the optimized waveguide crossing offers minimal -40 dB of the crosstalk in the C-band.

As the data reaches the aggregation unit, they have to be merged. Here again, interference between signals can be an issue. As discussed earlier in this subsection, we can make use of up to four modes without significant crosstalk between the signals. We can reuse the orthogonality of modes here again. Each subarray group can be assigned a mode using a mode converter (MC), before it merges with the waveguide carrying the signals to the aggregation unit's demultiplexer (demux). These changes to the architecture solve challenge (3).

*4) Addressing bit size mismatches*

OPCM cells within the photonic memory can be designed to have different bit densities, e.g., 1 bit/cell, 2 bit/cell, 4 bit/cell, etc. However, the parameters in an ML model like a CNN can be 32 bits in size without quantization. They can also be quantized to lower bitwidths such as 16 bits, 8 bits, or 4 bits to reduce storage requirements and to reduce computation latency and energy. In scenarios where there is a mismatch between OPCM cell bit density and the CNN parameter size (e.g., 4 bits/cell bit density with 8-bit CNN parameters), the one-shot multiplication operation achieved by reading the OPCM cell, as discussed earlier, is not feasible.

To support different bitwidth scenarios and tackle challenge (4), we make use of a time division multiplexing (TDM) based approach. For higher bit densities per cell than 4-bits (i.e., a nibble), each nibble will have to interact with every nibble of the other parameter. This can be achieved without significant loss in throughput because of solutions for challenges (1)-(3) which offer high parallelism in MAC operations, while the signals can stay disentangled from each other. However, we still have to perform shift-and-add operations to obtain the true results for these operations [39]. These necessary operations are facilitated within the aggregation unit (Fig. 5(b)). This results in an overall drop in throughput, but facilitates flexibility in operation, unconstrained by the OPCM MLC bit-density.

The aggregation unit is essential to address challenge (4), but it also provides some additional benefits. The photodetector (PD) based conversion to the electrical domain acts as a noise filtering mechanism. The wavelength-specific PDs offer disentanglement from crosstalk between wavelengths, improving SNR before the longer transmission to the E-O-E control unit. Additionally, the parameters can be stored within the SRAM cache within the aggregation unit, for additional accumulation operations if needed. We also consider 5-bit ADCs so that the data can be translated to the electrical domain with any carries from the operations. Finally, the readout signals for the MAC operations which were generated using low-power MDLs will be regenerated through DACs and vertical cavity surface emission lasers (VCSELs) for better fidelity before they reach the E-O-E controller which handles further aggregation and applies non-linear activation functions (see Fig. 3) for ML inference operations.

*D. CNN Mapping and Inference in OPIMA*

The architectural design choices discussed in the previous subsection allow the *OPIMA* architecture to realize high power consumption efficiency and high integrity large-scale parallel MAC operations and main memory accesses in the optical domain. From a CNN inference perspective, this offers two-fold benefits. Firstly, MAC operations are fundamental operations in CNNs and *OPIMA* can perform them with high degrees of parallelism. Secondly, CNNs in general require significant storage and data movement between layers, but this can be significantly reduced as the processing occurs within the memory where model parameters and activation feature maps are stored.

To leverage the parallelism offered by the PIM substrate in *OPIMA* for CNN inference, we need to efficiently map CNNs onto the OPCM arrays. For CNNs, this involves mapping the parameters from both convolutional layers and fully connected layers. Operations for both types of layers can be mapped into MVM operations. For convolutional layers, we adopt an input stationary dataflow approach, where the input data can stay in its native storage location while we drive the smaller weight matrices (decomposed as vectors) through them. Because of the large row sizes within the subarrays, we will be able to drive several kernels simultaneously. The feature map must be divided across subarrays, so that we can access subsequent rows of the map from neighboring subarrays. The kernels rows which must operate on the feature map can be encoded into laser signals through MDL tuning and be introduced into the subarrays. Additionally, we can achieve several parallel MAC operations through in-waveguide interference of WDM signals, from multiple subarrays within the same subarray group.

Let us consider a simple example with a 2×2 kernel, a feature map ($F$) with a row size of 4 elements, and MDL array generating wavelengths $\{\lambda_1, \lambda_2, ..., \lambda_C\}$ ($C$=number of columns per subarray). The kernel can be broken down into two vectors and mapped to MDL wavelengths: $k_1 = \{k_{00}, k_{01}\} \rightarrow \{\lambda_1, \lambda_2\}$ and $k_2 = \{k_{10}, k_{11}\} \rightarrow \{\lambda_1, \lambda_2\}$. Similarly the rows in $F$ can be broken down into vectors and mapped to subarrays: $\{f_{00}, f_{01}, f_{02}, f_{03}\} \rightarrow Subarray_1$ and $\{f_{10}, f_{11}, f_{12}, f_{13}\} \rightarrow Subarray_2$. Both subarrays must be within the same subarray group to facilitate the MAC operation. If we now enable access to the rows containing these vectors and simultaneously send the $k_1$ and $k_2$ signals from the MDLs through the subarrays, we shall obtain the following in the common readout waveguide bus $\{(k_{00} \times f_{00}, k_{10} \times f_{10}), \lambda_1\}, \{(k_{01} \times f_{01}, k_{11} \times f_{11}), \lambda_2\}$.

Because signals of the same $\lambda_i$ interfere with each other, this in turn generates: $(k_{00} \times f_{00} + k_{10} \times f_{10})$, $(k_{01} \times f_{01} + k_{11} \times f_{11})$, which is one addition away from generating the first element of an output feature map. This addition can be performed at the aggregation unit. The kernel can be moved across the MDL array to reflect the stride operation and further outputs can be obtained. Additionally, multiple kernels can be deployed simultaneously over $F$, across different wavelengths, reducing overall processing time requirement. This mapping process scales easily with kernel sizes as well, if the kernel sizes do not exceed the subarray row size.

For fully connected layers we opt for a weight-stationary approach. In both cases, the stationary matrix must be distributed across subarrays to ensure parallelism in operations. Once this mapping process is done, *OPIMA*'s PIM-specific architecture (Fig. 5), as described in this section, can be utilized effectively to achieve inference operation.

## V. EXPERIMENTS

In this section, we discuss the evaluation of the performance of *OPIMA* for PIM-based CNN inference acceleration. *OPIMA* adopts a main memory configuration of 4 banks, 64×64 subarrays per bank, with 256×512 OPCM elements and 256 MDLs per subarray. For evaluating *OPIMA* we rely on a modified NVMain 2.0 [61] for memory simulation followed by a Python-based performance analyzer, which makes use of the loss and energy parameters from detailed physics simulations and fabricated device characterizations summarized in Table 1.

We compare *OPIMA* against several electronic and optical acceleration platforms along with the current state-of-the-art photonic PIM. For photonic accelerator systems, we consider the work in [32], named PhPIM in our comparison studies, which proposed a PIM adjacent system, and CrossLight [41], a photonic CNN accelerator. CrossLight and PhPIM are modeled using the parameters in Table 1, and considering 8GB DDR5 DRAM, with 4800 megatransfers per second (MTS) data transfer rate as its main memory.

We also consider Nvidia P100 GPU (referred to as NP100 in results), AMD EPYC 7742 CPU (referred to as E7742 in results), and Nvidia Jetson ORIN (a low-power embedded GPU for edge AI applications; referred to as ORIN in results), as our electronic platform comparison points. Additionally, we consider the ReRAM based PIM CNN accelerator PRIME [11] for comparison.

TABLE I: OPTICAL LOSS AND POWER PARAMETERS CONSIDERED FOR *OPIMA*

| Loss parameters | Values |
| --- | --- |
| Directional coupler loss | 0.02 dB [42] |
| MR drop loss | 0.5 dB [43] |
| MR through loss | 0.02 dB [44] |
| Propagation loss | 0.1 dB/cm [45] |
| Bending loss | 0.01 dB/90º [46] |
| EO tuned MR drop loss | 1.6 dB [47] |
| EO tuned MR through loss | 0.33 dB [47] |
| SOA gain | 20 dB |

| Energy parameters | Values |
| --- | --- |
| OPCM read | 5 pJ [23] |
| OPCM write | 250 pJ [23] |
| EPCM write | 860 nJ [48] |
| DRAM access | 20 pJ/bit [49] |
| ADC | 24.4 fJ/step [50] |
| DAC | 2.0 pJ/bit [51] |

### A. Subarray grouping

The first experiment explores the *OPIMA* design space to determine the number of subarray groups, which in turn will determine the number of operations that can be performed per cycle, in *OPIMA*. This increase in parallelism trades off with the power consumption of the architecture. As the number of groups increases, the complexity of the interface required at the aggregation unit also increases, along with the laser power requirement to perform the operations. Simultaneously, we would like the maximum number of subarray rows to be accessible for main memory operations.

The *OPIMA* memory organization has 64 rows of subarrays per bank as mentioned earlier, which must be grouped as per the criteria discussed above. While considering the groups, we would like to avoid the extremes i.e., the case with a single group or the case with each subarray row belonging to an individual group, resulting in 64 groups. A single group severely limits parallelism, and 64 groups imply that all 64 rows will be engaged in PIM operations, essentially preventing any main memory read/write operations.

TABLE II: VARIOUS MODELS CONSIDERED FOR *OPIMA* EVALUATION AND THEIR ACCURACY ACROSS QUANTIZATION LEVELS FOR CLASSIFYING THE SPECIFIED DATASETS.

| Model | Dataset | Accuracy (fp32) | Accuracy (int8) | Accuracy (int4) | Parameter count |
|---|---|---|---|---|---|
| Resnet18 | CIFAR100 [57] | 75.3% | 74.2% | 72.6% | 11584865 (11.6 M) |
| InceptionV2 | SVHN [58] | 81.5% | 80.8% | 75.9% | 2661960 (2.6 M) |
| MobileNet | CIFAR10 [57] | 88.2% | 87.5% | 83.5% | 4209088 (4.2 M) |
| SqueezeNet | STL-10 [59] | 92.5% | 90.3% | 86.5% | 1159848 (1.1 M) |
| VGG16 | Imagenette [60] | 98.96% | 96.25% | 93.7% | 134268738 (134.3 M) |

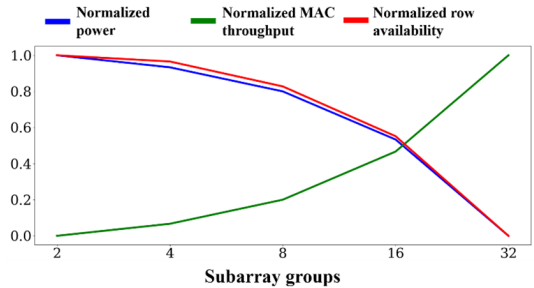

**Fig. 7:** Subarray group selection for *OPIMA* architecture.

Fig. 7 shows the normalized power, MAC throughput, and rows available for main memory operation, with changing number of subarray groups (x axis). It can be observed that a configuration with 16 groups strikes a balance between achievable compute parallelism with reasonable power consumption and sufficient memory access without starvation. Additionally, 16 subarray groups enable the maximum throughput efficiency (MAC/Watt) from *OPIMA*.

Our earlier analysis on mode conversion pointed to the fact that we can only have a maximum of four modes in our waveguide at the aggregation unit. Since we must rely on four modes only, to meet the demand of 16 groups, the modes can be reused. For enabling mode reuse, we use the same mode converter designs along the computation waveguides (Fig. 5(b)). Additionally, to prevent the same modes from interacting with each other, each of the four modes is assigned a separate multimode waveguide for transferring to the demux unit within the aggregation unit.

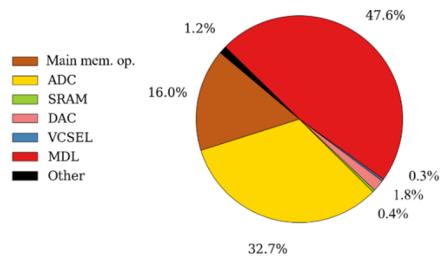

**Fig 8:** Power breakdown for *OPIMA* architecture.

### B. OPIMA power breakdown

The power consumption breakdown of the resulting version *OPIMA* is shown in Fig. 8. From this plot we can observe that the maximum power consumption is contributed by the MDL array and the electrical-optical interface, leading to a maximum power consumption of 55.9 W, for both main memory and PIM operations running simultaneously.

### C. CNN workload accuracy and latency analyses

For workloads we considered four CNN models: ResNet18 [53], InceptionV2 [54], MobileNet [55], and SqueezeNet [56]. The inference is performed for image classification of datasets, details of which are provided in Table II. We have considered 4-bit integer quantization using TensorRT, as this is the baseline MLC capacity. As the table shows this level of quantization results in at most 6% loss in accuracy, in the considered models. But this accuracy drop is model architecture-dependent, as can be seen in Table II. To showcase *OPIMA's* flexibility in handling parameter sizes, we have also considered 8-bit variants of the same models (Table II).

Before we go into further comparisons, we first analyze the performance of *OPIMA* using both the 4-bit and 8-bit quantized variants of the CNN models. A breakdown of *OPIMA's* latency in ms, as it processes these models, is provided in Fig. 9. Processing latency is the total time for processing the necessary MAC operations and the aggregation unit operation, i.e. all in-memory processing operations. The writeback latency refers to the latency incurred while applying the non-linearities and writing back the results, i.e. output feature maps, back into *OPIMA's* main memory architecture.

It can be observed that writeback is a significant contributor to latency in *OPIMA*. The PIM operations can leverage data within the memory and the high processing parallelism, leading to remarkably low processing times. However, the latency for the OPCM write operations needed to make the output feature maps available within the memory for further processing far outweighs the latency savings from the PIM operations. So, even though *OPIMA* can handle a variety of parameter sizes, given the OPCM write latencies, it is prudent to rely on 4-bit quantized models, while suffering some loss in accuracy, if throughput is significantly more important.

It can also be observed that *OPIMA* does not perform as one would expect for the far smaller InceptionV2 and MobileNet models when compared to ResNet18. Both models have higher processing latencies, with MobileNet having significantly higher processing latency than ResNet18. This is attributed to the 1×1 kernel in these models, which pose problems for the WDM-based MAC parallelization within *OPIMA*. Since the results from these operations do not have any further accumulation to be performed on them, they prevent the totality of the subarray row from being used. If more operations are performed, they will interfere with the results from the 1×1 kernel, leading to erroneous results. So, when these are encountered, *OPIMA* loses a significant portion of its parallel processing capabilities, especially when they are sequential in the CNN execution graph, like in the case of InceptionV2. MobileNet, though a larger model, offers higher parallelization opportunities, and hence performs at a similar latency, despite being ~4× the size of InceptionV2.

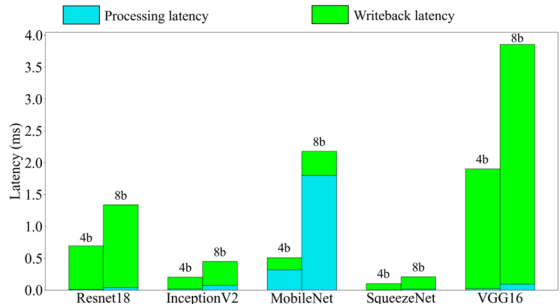

**Fig. 9:** Latency breakdown for *OPIMA*'s 4-bit (4b) and 8-bit (8b) variants across the models from Table II.

Similarly, writeback is a significant contributor to overall latency as discussed earlier. However, this is proportional to the sizes of the output feature maps generated by the model and not the computational complexity of the model. This is the reason MobileNet has lower writeback latency than processing latency, in comparison, and why InceptionV2 has an overall lower latency than ResNet18.

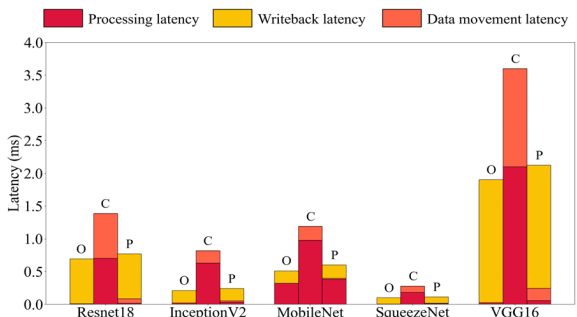

**Fig. 10:** Latency breakdown of CNN model inference across photonic architectures *OPIMA* (O), CrossLight (C), and PhPIM (P), for model-dataset pairs from Table II.

To further characterize the latency benefits of *OPIMA*, we compare it against the latency for the other photonic computing architectures we have considered, as shown in Fig. 10. The OPCM-based architectures (*OPIMA*, PhPIM) have better performance than CrossLight, because of the higher parallelism achievable in these architectures. PhPIM leverages the photonic tensor core operation from [15], along with an external DRAM acting as the actual main memory. PhPIM has opted for the faster yet energy-intensive electrical PCM programming mechanism, but the tensor core operation is still in the optical domain. The reprogramming, or writeback as we call it for an OPCM PIM, is significantly faster for PhPIM. However, *OPIMA* leverages much higher parallelism inherent to a main memory, and available to a PIM architecture, enabling faster processing times. Additionally, *OPIMA* does not have to access an external DRAM to access data needed for processing hence it does not have any external data movement latencies associated with its operation. Note that the internal data movement latencies are factored into our writeback latency.

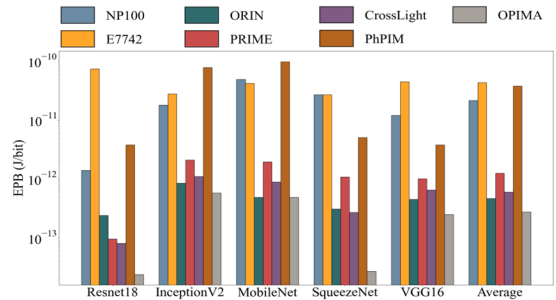

**Fig. 11:** EPB comparison across architectures

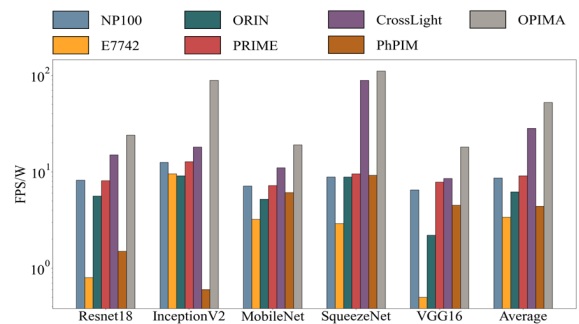

**Fig, 12:** FPS/W comparison across architectures

*D. Comparison studies*

In this section, we compare *OPIMA* against the various photonic and electronic acceleration platforms in terms of energy per bit (EPB) and throughput efficiency (FPS/W; FPS=frames per second).

On average *OPIMA* achieves 78.3×, 157.5×, 1.7×, 4.4×, 2.2× and 137× better performance in terms of EPB over NP100, E7742, ORIN, PRIME, CrossLight, and PhPIM respectively (Fig. 11). It should be noted that P100 can outperform *OPIMA* in terms of raw throughput, especially in the case of InceptionV2 and MobileNet, where the GPU threads are not constrained by the interference limitations of our WDM-based parallelization of operations. But *OPIMA* consumes significantly less power, which also leads to overall better throughput efficiency. In terms of FPS/W, *OPIMA* achieves 6.7×, 15.2×, 8.2×, 5.7×, 1.8×, and 11.9× better performance over NP100, E7742, ORIN, PRIME, CrossLight, and PhPIM respectively (Fig. 12).

It can also be noted that though *OPIMA* and PhPIM had comparable latencies (Fig. 10), *OPIMA* is able to outperform PhPIM in these metrics. This is because of the energy-intensive EPCM write processes that accompany PhPIM operation (nJ), as opposed to *OPIMA*'s OPCM reprogramming process (pJ).

## VI. CONCLUSIONS

In this work, we presented *OPIMA*, a high throughput, low latency, highly energy efficient OPCM PIM architecture. OPIMA showcases how an OPCM main memory architecture can be rearchitected to achieve photonic PIM. Through device-level design to enhance efficiency and various architectural innovations, *OPIMA* compares remarkably against electronic and photonic ML acceleration platforms. On average *OPIMA* outperforms the considered architectures by 83.1× in terms of EPB and 27.5× in terms of FPS/W. It outperforms the state-of-the-art photonic PIM architecture PhPIM by 186× and 55.3× in these metrics, while achieving lower average latency, across several CNN models. *OPIMA* also opens the door for possible system-level integration of photonic PIM with dedicated photonic accelerators such as those described in [20]-[22], [41]. Such a system can benefit from both the higher bandwidth that *OPIMA*'s main memory can provide along with computation support through PIM.